\documentclass[11pt]{article} 

\usepackage[utf8]{inputenc} 

\usepackage{geometry} 
\usepackage{graphicx} 
\usepackage{amsmath}
\usepackage{float}
    
\usepackage{booktabs} 
\usepackage{array} 
\usepackage{paralist} 
\usepackage{verbatim} 
\usepackage{subfig} 
\usepackage{multirow}
\usepackage{tabularx}
\usepackage{ltablex}
\usepackage{authblk}

\usepackage{fancyhdr} 
\providecommand{\keywords}[1]
{
  \small	
  \textbf{\textit{Keywords---}} #1
}

\usepackage{sectsty}
\allsectionsfont{\sffamily\mdseries\upshape} 

\usepackage[nottoc,notlof,notlot]{tocbibind} 
\usepackage[titles,subfigure]{tocloft} 

\title{Forecasting local hospital bed demand for COVID-19 using on-request simulations}

\author[1]{Angelo D'Ambrosio}
\author[1]{Raisa Kociurzynski}
\author[1]{Alexis Papathanassopoulos}
\author[1]{Fabian Bürkin}
\author[1]{Hajo Grundmann}
\author[1]{Tjibbe Donker\footnote{Corrispondence to: tjibbe.donker@uniklinik-freiburg.de}}

\affil[1]{\footnotesize Institute for Infection Prevention and Hospital Hygiene, Freiburg University Hospital, Freiburg, Germany.}
\begin{document}

\maketitle

\abstract{
For hospitals, realistic forecasting of bed demand during impending epidemics of infectious diseases is essential to avoid being overwhelmed by a potential sudden increase in the number of admitted patients.
Short-term forecasting can aid hospitals in adjusting their planning and freeing up beds in time.

We created an easy-to-use online on-request tool based on local data to forecast COVID-19 bed demand for individual hospitals.
The tool is flexible and adaptable to different settings, and it is based on a stochastic compartmental model for estimating the epidemic dynamics and coupled with an exponential smoothing model for forecasting.

The models are written in R and Julia and implemented as an R-shiny dashboard.
The model is parameterized using COVID-19 incidence, vaccination, and bed occupancy data at customizable geographical resolutions, loaded from official online sources or uploaded manually.
Users can select their hospital's catchment area and adjust the number of COVID-19 occupied beds at the start of the simulation.
The tool provides short-term forecasts of disease incidence and past and forecasted estimation of the epidemic reproductive number at the chosen geographical level.
These quantities are then used to estimate the bed occupancy in both general wards and intensive care unit beds.
The platform has proven efficient, providing results within seconds while coping with many concurrent users.

By providing ad-hoc, local data informed forecasts, this platform allows decision-makers to evaluate realistic scenarios for allocating scarce resources, such as ICU beds, at various geographic levels.
}


\section{Introduction}
The speed of community transmission of virulent pathogens during epidemics or pandemics of infectious diseases has the potential to overwhelm healthcare systems \cite{li_demand_2020}.
Hospital beds are usually in limited supply, because most hospitals run at near-full capacity under normal circumstances.
A sudden increase in patients needing hospital care might thus not be met with the required available beds if the growth in demand is unexpected.
Reliable forecasting of bed demand during epidemics is therefore essential for hospitals to free up beds in time.

However, epidemic trajectories can differ considerably regionally \cite{white_state-level_2020},
and because patients are usually admitted to hospitals in their own region \cite{Donker2013},
the bed demand can widely vary between hospitals as well.
Consequently, nationwide forecasts of bed demand may not reflect the individual hospital's situation, and may fail to predict overload problems.
We, therefore, developed a bed demand forecasting model \cite{donker_navigating_2021} based on local epidemic trajectories in the hospital's individual catchment area and its current bed occupancy.

To be widely applicable, this model needs to be able to produce bespoke forecasts for each hospital in a country for which basic levels of data are available.
This capability requires a flexible forecasting system because of the vast number of potential combinations of parameter choices for all individual hospitals,
in particular concerning the possible choices for catchment areas per hospital, the number of currently occupied beds, and patients' lengths of stay on various wards.
Such a variety of parameters enables the extreme forecast flexibility required to produce ad-hoc reports and forecasts for all situations of specific hospitals.

We made the use of this model accessible for non-technical users by creating an online interactive dashboard.
This platform allows individual users (i.e., hospital managers) to enter the data and parameters related to their specific context and requirements and produce ad-hoc bed occupancy estimates and predictions.
End user-specific forecasting of bed demand is a unique concept, as on-request forecasting is often avoided because of the computational challenges of multiple concurrent user modelling platforms.

Here we show how some of these computational challenges can be solved. 
Careful streamlining of the model and its technical implementation can help produce epidemic forecasts in a reasonable time,
allowing users to explore the potential future epidemic trajectory without being restricted to the developer's viewpoint.

We will first discuss the forecast model itself, its data and parameter requirements, as well as the different modules it contains. 
Each module is designed to guide the user through the steps and assumptions needed to produce the forecast.
Then we will discuss the technical implementation of the online dashboard and the technologies that ensure usability for multiple concurrent users.
Finally, we will present performance statistics of the dashboard collected between December 2021 and March 2022.

The code used to implement the dashboard, in the version used to prepare this manuscript, is available at: https://github.com/QUPI-IUK/Bed-demand-forecast/releases/tag/v.0.5.6.

\section{Methods}
\subsection{General model structure}
\begin{figure}[htb]
\begin{center}
\includegraphics[width=\textwidth]{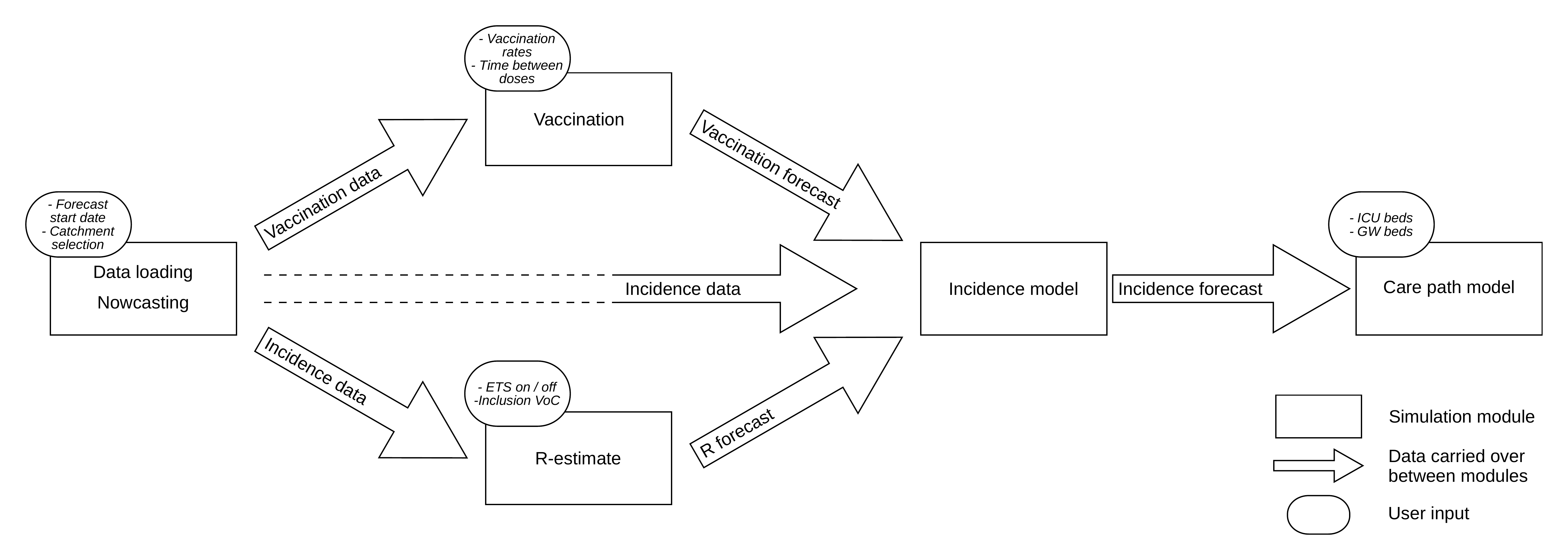}
\caption{Model structure showing each of the modules, the data carried over between modules, and the possible user input in each module.}
\label{fig:schematic}
\end{center}
\end{figure}

The complete forecast model consists of 5 main modules: 1) Data loading and nowcasting, 2) Reproduction number estimates, 3) Vaccination coverage forecasting, 4) Incidence model, and 5) Care path model.
The calculations done in the first two modules rely only on the loaded data and user input,
while modules 3-5 also build further on the results of each previous module
(see figure \ref{fig:schematic}).
This hierarchical structure implies that any change to parameters in a module updates the outputs in all dependent modules.
However, by guiding the user through the modules in sequential order, the number of required computations can be drastically reduced.

\subsection{Data structure and definitions}
The model needs data on infection incidence, vaccine doses, bed occupancy in general wards and intensive care units, and a simulation window for which a forecast is required. 
Although we developed this model for hospitals in Germany, based on data provided by the Robert Koch Institute \cite{rki_data} and the Deutsche Interdisziplinäre Vereinigung für Intensiv- und Notfallmedizin (DIVI) \cite{divi_data}, the model can be applied to any country that has the following data available (preferably if in remote-accessible, machine-readable format):
\begin{enumerate}
\item incidence, $I_{g}(t)$
\item administered vaccinations, $V_{d,g}(t)$
\item bed occupancy on general wards $B_{GW,g}(t)$ and ICU $B_{ICU,g}(t)$
\end{enumerate}
where $t$ denotes the time in days since a reference date before the start of the pandemic, $g$ denotes a geographical subdivision (in the case of Germany, this is a district, i.e., Landkreis or Stadtkreis), and $d$ denotes the dose of vaccine administered (1st, 2nd, or 3rd/booster). 
Incidence and vaccination data are needed for each day of the period or reference,
while merely the value at the start date of the forecast is needed for bed occupancy.

The user defines the catchment area of interest, selecting geographical areas from the available list
($K_{all}$), 
into the catchment set ($K_u$). 
After this, the input data is summed over the selected areas: 
$$I(t) = \sum_{g\in K_u} I_g(t),$$
$$V_d(t) = \sum_{g\in K_u} V_{d,g}(t),$$
$$B_{GW}(t) = \sum_{g\in K_u} B_{GW,g}(t),$$
$$B_{ICU}(t) = \sum_{g\in K_u} B_{ICU,g}(t).$$

The start of the simulation is denoted as $T_S$, 
for which the incidence, number of vaccinations, and bed occupancy are taken from the (last data-point of the) data, the first forecasted day is thus $T_S+1$. 
The length of the forecast is defined as $L$; the last forecasted date is therefore given as $T_E = T_S+L$.
The size of any class of individuals in the model is given in absolute numbers of individuals; thus $S(0) = N$, with $N$ being the population size.

Because many of the parameters in the dashboard's models are continuously distributed (as Exponential, Gamma, or Weibull),
and the model uses discrete time steps,
we need to discretise the parameter distributions.
For each continuous distribution, with probability density function (PDF) $f_c(t)$, the discretised PDF is defined as $f_d(t) = \int_t^{t+1}f_c(x)dx.$

User input for all distributions is given in terms of the mean $\mu$ and standard deviation $\sigma$, and distribution parameters are then determined by moment matching. 
For a gamma distribution, this means that 
$f(x,\alpha,\beta) = f(x,\alpha=\mu^2/\sigma^2, \beta=\mu/\sigma^2)$, 
for a Weibull distribution, 
$f(x, k, \lambda) = f(x,k=(\frac{\sigma}{\mu})^{-1.086}, \lambda=\mu / \Gamma(1+1/k))$, 
and for an exponential distribution 
$f(x,\lambda)= f(x, 1/\mu).$

\subsection{Vaccination model}
The vaccination model consists of two parts. 
In the first, we forecast the number of people vaccinated over the length of the forecast for each of the doses (1st, 2nd, and 3rd/booster dose), 
while in the second part we convert these administered vaccinations into population-level protection against transmission.

The first doses ($V_{1}(t)$) are assumed to be administered at the same rate as during the last observed week: $$V_{1}(t>T_{S}) = \sum_{i=T_S-6}^{T_S}V_{1}(i)/7$$
The second dose is assumed to be administered at a fixed time delay after the first dose. This time delay $\Delta T_V$ is extracted from the data as the maximum time for which 
$$V_{1}(T_S-\Delta T_V) \geq V_{2}(T_S) $$
holds.
The second doses are then a direct reflection of the number of first doses $\Delta T_V$ days ago,
$$ V_{2}(t) = V_{1}(t-\Delta T_V).$$

The daily administered third/booster doses are assumed to be a continuation of the mean daily third doses over the previous week: 
$$V_{3}(t>T_{S}) = \sum_{i=T_S-6}^{T_S}V_{3}(i)/7.$$
This is in line with the forecasting of the first doses, with the exception that the cumulative number of third doses is not allowed to exceed the cumulative number of second doses a certain delay ($\Delta T_B$) ago, such that
$$\sum_{i=0}^{T_S} V_{3}(i) \leq \sum_{i=0}^{T_S-\Delta T_B} V_{2}(i)$$
has to hold.

\subsubsection{Population protection}
Each dose is assumed to have a time-dependent additive effect on the population-wide protection against infection. 
At the individual level, $G_{d}(\tau)$ denotes the proportion of individuals no longer susceptible to infection $\tau$ days after the administration of dose $d$.
This function thus serves to simulate the delay between vaccine administration and maximum vaccine protection at the individual level.
We assume no waning of immunity, such that $G_d(\infty) = E_d$, with $E_d$ the vaccine effectiveness against transmission elicited by doses $d$.

The additive effect on vaccine effectiveness of each additional dose is defined as ${E_d}^*= E_d-E_{d-1}$;
\footnote{Consequently, ${E_2}^*= E_2-E_{1}$ and ${E_3}^*= E_3-E_{2} = E_3-({E_{2}^*}+{E_{1}})$}.
consequently, the additive increase in an individual's protection due to an extra dose can be written as:
$${G_d}^*(\tau) = G_d(\tau) \frac{{E_d}^*}{E_d}.$$

$G_d(\tau)$ is input as the cumulative distribution function (CDF) of a normal distribution with mean $\mu(G_1)=15$, $\mu(G_2)=15$, $\mu(G_3)=7$, and standard deviation $\sigma(G_1)=3.8$, $\sigma(G_2)=6.5$, $\sigma(G_3)=3.8$ as default values. 

The population protection at time $t$ is then given by,
$$G_p(t) = \sum_{d}^{\{1,2,3\}}\sum_{i=0}^{t-1} \left( {G_d}^*(t-i) \frac{V_{d}(i)}{N} \right).$$
which is used in both the R forecasting step and the incidence model.

\subsection{R estimation and forecasting}
In order to inform the transmission process part of the incidence model, 
we need to forecast the reproduction number of the pathogen over the model timeframe. 
This forecast is based on the observed time-varying effective reproduction number ($R_e(t)$), 
and implemented either as the static continuation of the most recent observation, 
or forecasted using an ETS (Error, Trend, Seasonal) model \cite{Hyndman_forecasting_2021} based on the last 100 days of observations, depending on the user's preference.
Furthermore, we provide the option to incorporate the effect of a Variant of Concern (VoC) on the development of the $R_0(t)$. 
In all cases, we use estimated values of $R_e(t)$ from the available data up to the simulation start date $T_S$, $R_e(t\leq T_S)$, and forecasted values after the simulation start date ($R_e(t>T_S)$).

We estimate the time-varying effective reproduction number $R_e(t)$
using the Cori et al. R estimation method \cite{cori_new_2013} as implemented in the EpiEstim R package (version 2.2-4), with a given non-parametric Serial Interval (SI).
This SI distribution denotes the number of days between equal disease events (e.g., onset of symptoms) of directly connected cases. 
The default SI is assumed to be Gamma distributed with a mean of 5 days and a standard deviation of 4.9.
These values were chosen to reflect estimates of mean SI from multiple studies \cite{lehtinen_relationship_2021} with a relatively high standard deviation.

By keeping track of the cumulative number of infected individuals, as well as the previously calculated population protection ($G_P(t)$), we can calculate $R_0(t)$ from $R_e(t)$, 
given that $R_e$ is related to $R_0$ through the susceptible population:
$$S(t) = N (1 - \sum_i^t \frac{I(i)}{N}) (1-G_P(t))$$
$$R_e(t) = R_0(t) \frac{S(t)}{N},$$
$$\Rightarrow R_0(t) = \frac{R_e(t) N}{S(t)}.$$

\subsubsection{ETS model}
To produce a dynamic forecast of $R_0(t)$, we used an ETS model which underlies an exponential smoothing method \cite{Hyndman_forecasting_2021}.
In general, exponential smoothing methods generate point forecasts of time series using weighted averages.

For an observed time series $y_1, y_2,...,y_n$ the point forecast for $h$ periods ahead is $y_{t+h}$. When the forecast is based on all data up to time $t$, it is denoted as $y^*_{t+h|t}$. For a method with an additive trend, the point forecast results from the level of the series $l$ and the growth $b$ at time $t$:

$$y^*_{t+h|t} = l_t + h b_t $$

which gives the forecast equation.

While $l_t$ and $b_t$ are based on the level and trend (slope) at $t-1$ and are given by the smoothing equations:

$$l_t = \alpha y_t + (1-\alpha)(l_{t-1} + b_{t-1}),$$
$$b_t = \beta^*(l_t - l_{t-1}) + (1 - \beta^*)b_{t-1}.$$

This approach is also known as Holt's linear method. Values for the initial states $l_0$ and $b_0$ as well as the smoothing parameters $\alpha$ and $\beta^*$ are estimated from the observed data, with $0 < \alpha, \beta^* < 1$. Increasing $\alpha$ gives more weight to the more recent observations and less weight to observations that lie longer ago. In our case, $\alpha$ and $\beta^*$ were set to 0.25 and 0.15, respectively.

While the exponential smoothing methods generate point forecasts, the underlying statistical models, in addition, generate prediction intervals. To generate such an ETS forecast model, we need to specify the probability distribution for the residual at time $t$, $e_t$. For additive errors, we assume the residuals are normally and independently distributed with mean 0 and variance $\sigma^2$, notated as $e_t = \varepsilon_t \sim$ NID$(0,\sigma^2)$.
Expressing the error as $\varepsilon_t = y_t - l_{t-1} - b_{t-1} \sim$ NID$(0,\sigma^2)$ we can substitute $\varepsilon_t$ into Holt's linear method

$$y_t = l_{t-1} + b_{t-1} + \varepsilon_t,$$
$$l_t = l_{t-1} + b_{t-1} + \alpha \varepsilon_t,$$
$$b_t = b_{t-1} + \beta \varepsilon_t.$$

with $\beta = \alpha \beta^*$. These expressions are referred to as the state-space models as they include an equation for the observation and equations for the unobserved states (level, trend). Each observation $y_t$ in the ETS model therefore includes a component for the level (or smoothed value), 
an error component (E), a trend component (T), and seasonality (S). In our case, we use the ETS(A,A,N) model, which has additive errors, additive trend and no seasonality;

We use a $\log(R_0(t))$ time series of a 100 days, i.e. $\log(R_0(T_S-99 ... T_S))$, as input for the ETS model. 
The logarithm serves to forecast the relative changes in $R_0$ and avoids negative $R_0$ forecasts.
Per model iteration, we pick a random quantile for the prediction interval and use this as the trajectory for $R_0(t>T_S)$. 

\subsubsection{Variant of Concern implementation}
To model the impact of new pathogen variants emerging in the population \cite{abdool_karim_new_2021}, we need to consider how a variant of concern (VoC) has an advantage over the background viral population.
We considered two possible mechanisms: 
1) through higher transmissibility, increasing the base $R_0(t)$, 
and 2) by immune evasion, effectively increasing the size of the susceptible class ($S(t)$).

In case of increased transmissibility, we increase the basic reproduction number of the background viral population $R_{0,B}(t)$ by the relative added fitness advantage $A$, 
$$R_{0,V}(t) = (1+A)R_{0,B}(t).$$

Note that the previously calculated $R_0(t)$ is the average of the background and VoC reproduction numbers weighted by the proportion of cases caused by either 
(respectively $\rho_B(t)$ and $\rho_V(t)$), 
\begin{align*}
R_0(t) &= \rho_B(t) R_{0,B}(t) + \rho_V(t) R_{0,V}(t),\\
 &= (1 - \rho_V(t)) R_{0,B}(t) + \rho_V(t) (1+A) R_{0,B}(t).
\end{align*}

We can calculate $R_{0,B}(t)$, used as the general reference reproduction number in the model, from $R_0(t)$ as
$$R_{0,B}(t) = \frac{R_0(t)}{(1-\rho_V(t))+\rho_V(t)(1+A)},$$ 

or we can directly calculate this from the observed $R_e(t)$ value 
$$R_{0,B}(t) = \frac{R_e(t) N }{\left((1-\rho_V(t))+\rho_V(t)(1+A)\right)S(t)}.$$

The proportion of cases caused by the VoC ($\rho_V(t)$) over time can be described with a sigmoidal function. 
A similar approach to estimating fitness advantages for new variants has been used by others \cite{althaus_tale_2021}.
This is done by assuming both the background population and the VoC have a static growth rate unaltered by added immunity through new cases
(i.e., the relation between their growth rates is fixed over time).
We can calculate these growth rates ($r$) as a function of $R_0$, using the mean serial interval ($\sigma$) \cite{Wallinga_2007}

$$r=\frac{\log(R_0)}{\sigma}.$$ 
This means that the number of infections from both background variants and the VoC can be described as respectively 
$$I_B(t) = I_B(0) e^{\frac{\log\left(R_{0,B}\right)}{\sigma}t}$$ 
and 
$$I_V(t)=I_V(0) e^{\frac{\log\left((1+A)R_{0,B}\right)}{\sigma}t}.$$

The proportion VoC is
$$\rho_V(t) = \frac{I_V(t)}{I_V(t)+I_B(t)}= \frac{1}{1+\frac{I_B(t)}{I_V(t}},$$ 
which converts to 
$$\rho_V(t) = \frac{1}{1+\frac{I_B(0) e^{\frac{\log\left(R_{0,B}\right)}{\sigma}t}}{I_V(0) e^{\frac{\log\left((1+A)R_{0,B}\right)}{\sigma}t}}}.$$
By defining the total number of infected individuals as $I_T(0) = I_V(0)+I_B(0)$, and using $I_B(0) = (1-\rho_V(0)) I_T(0)$, we get 
$$\rho_V(t) = \frac{1}{1+\frac{1-\rho_V(0)}{\rho_V(0)}\frac{e^{\log\left(R_{0,B}\right) t/\sigma}}{e^{\log\left((1+A)R_{0,B}\right)t/\sigma}}},$$ 
rewriting to 
$$\rho_V(t) = \frac{1}{1+\frac{1-\rho_V(0)}{\rho_V(0)}e^{-\frac{\log(1+A)}{\sigma}t}}.$$ 

This shows that we only need the relative added fitness advantage ($A$) in combination with a starting proportion of the variant ($\rho_V(t=t_\rho)$) at a certain reference date ($t_\rho$) to describe the proportion VoC over time.
In the dashboard code, the reference date for the proportion of VoC does not overlap with the starting date of the incidence, vaccination, and bed occupancy data. 
In other words, with $I(t_I=0)$ and $\rho_V(t_\rho=0)$, $t_I\neq t_\rho$, because $\rho_V(t_I=0) = 0$, making a precise estimate of $\rho_V(t)$ impossible, or at least impractical. 
We, therefore, set $t_\rho$ to a later time point at which the $\rho_V$ is identifiable.

The added relative fitness advantage of the VoC can then be estimated from data about the proportion VoC among all isolates per week in the given catchment area, available through for instance genomic surveillance efforts.
The user can then set the reference date, starting proportion, and added relative fitness advantage based on such analyses to be used in the forecast. 

\subsubsection{Immune evasion}

In case the VoC has an advantage over the background variant through (partial) immune evasion, the number of individuals susceptible to the VoC ($S_V(t)$) would be greater than for the background variant ($S_B(t)$), giving it a higher $R_e(t)$ even if the $R_0(t)$ would remain the same. 

We define immune evasion ($\epsilon$) as the proportion of individuals immune against infection with the background variant that are not immune against infection with the VoC.
Thus, the number of individuals immune to infection with the VoC scales linearly with the number of individuals immune to infection with the background variant, if very few or no infections with the VoC have occured yet.

To correctly assess the added fitness advantage at the VoC reference point ($t_\rho$), we need to calculate the increase in the number of individuals susceptible to infection with the VoC relative to those susceptible to the background variant.
For simplicity, we assume that the cumulative number of infections with the VoC is negligible relative to the cumulative number of infections with the background variants at $t_\rho$, that vaccination is aimed at the background variants, and immunity against the VoC is only determined by cross-immunity ($1-\epsilon$) from immunity against the background variants.

$$S_B(t_\rho=0) = N (1-G_P(t_\rho))\left(1- \sum_{i=0}^{t_\rho}\frac{ I(i)}{N}\right) $$
$$S_V(t_\rho=0) = N-(1-\epsilon)(N-S_B(t_\rho=0))$$

Given that $R_{e,B}(t) = R_0(t) S_B(t)/N$ and $R_{e,V}(t) = R_0(t) (1+A) S_V(t)/N$, 
the relative added advantage by immune evasion, given the pre-existing immunity against the background variant, is
$$A_\epsilon = \frac{S_V(t_\rho)}{S_B(t_\rho)}-1,$$
and the total added advantage is
\begin{align*}
A^*&= (1+A)(1+A_\epsilon)-1,\\
&= (1+A)\frac{S_V(t_\rho)}{S_B(t_\rho)}-1.
\end{align*}

As $A^*$ is also the observed fitness advantage, the user will need to make an assumption about what part of this advantage is caused by immune evasion or by increased transmissibility on the reference date.
Although the reference date ($t_\rho$) and the start of the simulation ($T_S$) are not necessarily the same,
since the user can define $t_\rho$ independently of $T_S$, 
from here on, we only consider the case where $T_S=t_\rho$, for clarity and brevity. 


\subsection{Incidence model}
The incidence forecast model is based on a stochastic implementation of a Susceptible-Infected-Recovered (SIR) model.
The model describes the number of new cases as a function of the population still susceptible ($S(t)$) to infection and of infection pressure ($P(t)$) exerted by all those currently infectious, and the total population size ($N$).
The size of the components is given as absolute numbers of individuals; thus $S(0) = N$.
The infection process is governed by the forecasted basic reproduction number $R_0(t\geq T_S)$ computed in the preceding modules of the framework.

If no VoC is defined, the model describes the dynamics of a single variant (the background variant).
Conversely, the model converts into a two-strain model if a VoC is defined, keeping track of the infected individuals for either variant as well as the total number of individuals susceptible to either variant. 

\subsubsection{Single variant model}

The size of the susceptible class is determined by both the cumulative number of infected individuals and the population level protection through vaccination, as defined before, 
$$S(t) = N (1-G_P(t))\left(1- \sum_{i=0}^{t}\frac{ I(i)}{N}\right).$$
The infection pressure exerted on $S(t)$ is then given as the weighted number of infected individuals in the preceding serial interval, 
$$P(t) = \sum_{i=0}^t H(i)I(t-i) $$
These are thus all past cases that are able to cause new cases on the current day $t$, weighted by the probability that they do so $i$ days after they were reported themselves.
Given the current $R_0(t)$ and $P(t)$, the expected mean total number of newly infected individuals on day t+1 is  
$\overline{I}(t+1)=R_0(t) P(t)$. 

The transmission rate per infected individual 
(The probability of infecting each of the other individuals in the population in a unit of time), 
is given as $\beta(t) = R_0(t)/\delta N$. 
Since we use infection pressure $P(t)$ calculated over the entire SI, we can set $\delta$ to 1, such that $\beta(t) = R_0(t)/N$.

The infection pressure exerted on each individual within the population is then calculated as 
$$ P_I(t)= 1- (1-\beta(t))^{P(t)},$$
which for low numbers of infected individuals (and therefore low $P(t)$) can be simplified to $ P_I(t)= \beta P(t).$
The newly infected individuals ($I(t+1)$) are then randomly picked from a binomial distribution $I(t+1)=Binom(p=P_I(t),N=S(t))$.

\subsubsection{VoC implementation}
If a VoC is defined, we need to track the size of both the susceptible and the infected class of individuals for both the background variant and the VoC. 
This is first done retrospectively, for $t\leq T_S$, to determine the size of the classes at $T_S$.

With an estimate of the VoC proportion over time ($\rho_V(t\leq T_S)$), the past infections caused by the VoC can now be calculated as
$$I_V(t\leq T_S)=\lfloor \rho_V(t)I(t) \rceil,$$
where $\lfloor ... \rceil$ denotes rounding to the nearest integer. 
The number of infections with the background variants thus follows as $I_B(t) = I(t)- I_V(t).$ 
This also gives us the exact size of the susceptible class for both the background variant and the VoC

$$S_B(t) = N \left((1-G_P(t))\left(1- \sum_{i=0}^{t}\frac{I_B(i)}{N}\right)\right) \left(1-(1- \epsilon)\sum_{i=0}^{t}\frac{ I_V(i)}{N} \right),$$

$$S_V(t) = N\left((1-(1-\epsilon)G_P(t))\left(1- (1-\epsilon)\sum_{i=0}^{t}\frac{I_B(i)}{N}\right)\right)\left(1- \sum_{i=0}^{t}\frac{ I_V(i)}{N}\right).$$

Note that in the absence of new cases, the susceptible classes still need to be calculated forward in time before the incidence model is actually run, as forward vaccinations reduce their size.

The forward, prospective, model of the infection process, for $t>T_S$, is then performed for both the background variants and VoC in the same way as with a single variant, splitting $I(t)$, $P(t)$, $P_I(t)$, and $S(t)$ in a background variant ($*_B$) and VoC ($*_V$) class.
The final epicurve then combines both variants again, 
$$I(t)=I_B(t)+I_V(t).$$

\subsection{Care path (Within-hospital) model}
To forecast bed occupancy in the hospital, the main endpoint of this work, we first simulate the path of each single patient admitted to the hospital through the general ward, ICU, and step-down units (care path model), as described in Donker et al (\cite{donker_navigating_2021}). 
Then we join this model with the predicted incidence to forecast the hospital admission rate and the bed occupancy.

We use the following parameters to model the care path of individual patients through the hospital.

\begin{itemize}
\item Length of stay (LoS) distribution on general wards (GW)
\item LoS distribution on intensive care units (ICU)
\item LoS distribution on step-down units (SDU)
\item proportion of patients being transferred from GW to ICU
\item proportion of patients being transferred from ICU to SDU
\end{itemize}
It is possible to add the proportion of patients directly admitted to ICU ($H_I$) to this list, but this is usually estimated from the data (see below).

Once all admissions and discharges are determined, we track the total number of patients on each ward type on each day of the forecast, presenting the final result of the forecast.
Note that, despite modelling the general ward and step-down unit separately, we combine both wards into a single number of beds occupied on the general ward in any other step, because the occupied beds in both cases are usually reported as part of the same type (i.e., occupied non-ICU beds). 

\subsubsection{Admission rate}
To estimate the proportion of reported cases being admitted to hospital, we need to compare the incidence of reported cases to the total number of patients in hospital at a given moment (prevalence).
This can be achieved by taking the length of stay of admitted patients into account. 
We create a complete Length of Stay distribution $L(t_a)$, 
defined as the proportion of patients still in hospital $t_a$ days after admission
by simulating the care paths of a large number of patients ($N_P=10000$) with the same admission date using these parameters.
For each day after admission, we then track how many patients are still in hospital ($N_H(t_a)$), resulting in the Length of Stay distribution, 
$$L(t_a)=\frac{N_H(t_a)}{N_P}.$$

The admission rate can then be directly estimated using the current number of occupied beds in general wards $B_{GW}$ and in ICU $B_{ICU}$
$$\alpha (t) =\frac{B_{GW}(t)+B_{ICU}(t)}{ \sum_{i=0}^{t} I(i) L(t-i) },$$

which we assume to remain stable after the start of the simulation: $\alpha (t>T_S) = \alpha (T_S)$).

The same method delivers the distribution of admitted patients on the GW $L_{GW}(t_a)$ and on the ICU $L_{ICU}(t_a)$.
Note that $L(t_a) = L_{GW}(t_a)+L_{ICU}(t_a)$,
and while $L(0)=1$, $L_{GW}(0) \leq 1$ and $L_{ICU}(0) \leq 1$. 
To be exact, $L_{ICU}(0) = H_I$, reflecting the directly to ICU admitted patients, and $L_{GW}(0) = 1- H_I$,

The admission rate can also be calculated for the separate wards,
$$\alpha_{GW}(t) =\frac{B_{GW}(t)}{ \sum_{i=0}^{t} I(i) L_{GW}(t-i) },$$
$$\alpha_{ICU}(t) =\frac{B_{ICU}(t)}{ \sum_{i=0}^{t} I(i) L_{ICU}(t-i) }.$$

We can calculate the expected number of ICU beds under the assumption that no patients enter the ICU directly.
$${B^*_{ICU}}(t)=\left(B_{GW}(t)+B_{ICU}(t)\right)\frac{ \sum_{i=0}^{t} I(i) L_{ICU}(t-i)}{\sum_{i=0}^{t} I(i) L(t-i) }.$$
The surplus of ICU patients ($B_{ICU}(t)-{B^*_{ICU}}(t)$) is then caused by the direct admission of patients to ICU, and calculated as the ICU surplus over the total number of occupied beds.

$$H_I(t) = \frac{B_{ICU}(t)-{B^*_{ICU}}(t)}{B_{GW}(t)+B_{ICU}(t)}$$

Similar to $\alpha(t)$, we assume that $\alpha_{GW}(t)$, $\alpha_{ICU}(t)$, and $H_I(t)$ remain stable after $T_S$.

\subsubsection{Admission rate and VoC}
In general, the assumption that the admission rate is stable over time reasonably holds unless the dynamics of the disease suddenly change.
For now, we consider one such case: the occurrence of a variant with significant immune evasion, causing infections in individuals with immunity against the background variants through infection or vaccination 
(here called "reinfections", but it includes breakthrough infections to a large extent).
In such a case, immunity in these individuals may protect against severe disease (to some degree).
A sudden increase in the proportion of infections in these exposed individuals may lower the admission rate drastically.

We first need to calculate the number of reinfections.
To do this, we track the proportion of individuals that have any immunity against any or either variant, assuming the second dose of the vaccine gives enough protection to reduce hospitalisation rates significantly,

$$U(t)=1-\left( 1-\sum_{i=0}^{T_S}\frac{I(i)}{N}\right) \left(1-\sum_{i=0}^{T_S}\frac{V_2(i)}{N}\right),$$
$$U_B(t)=1-\left( 1-\sum_{i=0}^{T_S}\frac{I_B(i)}{N}\right) \left(1-\sum_{i=0}^{T_S}\frac{V_2(i)}{N}\right),$$
$$U_V(t)=\sum_{i=0}^{T_S}\frac{I_V(i)}{N}.$$

The number of reinfections is then determined by comparing the susceptible class ($S$) with the previous immunity class ($U$).
The sizes of these classes differ because of non-complete protection from vaccination and cross-immunity between strains.

$$X_B(t) = \frac{S_B(t) - (1-U_B(t))(1-U_V(t))N}{S_B(t)}I_B(t),$$
$$X_V(t) = \frac{S_V(t) - (1-U_B(t))(1-U_V(t))N}{S_V(t)}I_V(t).$$

The proportion of reinfections among all infections is thus $${X^*}(t) = \frac{X_B(t)+X_V(t)}{I(t)}$$

The admission rate can then be adjusted to consider the effect of reinfections.
This is done by adjusting future admission rates relative to the observed admission rate at the start of the simulation.
First, we calculate the risk of admission for immunologically naive individuals (the "Primary hospitalisation risk"), $\pi$, at the start of the simulation 
$$\pi = \frac{\alpha(T_S)}{(1-{X^*}(T_S))+(\gamma {X^*}(T_S))},$$ 
given the relative admission risk for secondary infections $\gamma$.
The admission rate over time then develops with ${X^*}(t)$
$$\alpha^*(t) = {X^*}(t) \pi \gamma + (1-{X^*}(t))\pi $$

This admission rate then determines the number of individuals admitted to the hospital, entering the care path model, pulled from the binomial distribution 
$$Binom(p=\alpha^*(t),N=I(t)).$$

The care path model then depends on determining lengths of stays on each ward and movements between ward types for each individual admitted patient sampled from their respective distributions. 

However, the care path model needs to determine the lengths of stays for the patients already in hospital at $T_S$, that is $B_{GW}(T_S)$ and $B_{ICU}(T_S)$. 
To simulate their future discharge and transfer events, we create an admission record using a uniform number of patients per day for the 100 days preceding $T_S$ and simulate their care paths.

We then select all patients present in the GW or ICU at $T_S$, creating the "Current" patient population in the hospital. 
From the current population, we randomly select $N_{GW} = B_{GW}(T_S)$ and $N_{ICU} = B_{ICU}(T_S)$ patients to recreate the discharges for the current population.

\subsection{Dashboard server implementation}
The main backbone of the model is written in R (version 4.1.2 \cite{theRproject}) as an R-Shiny \cite{shiny} (version 1.6) app running using shiny-server on an 8-core, 16GB ram, Ubuntu (version 20.04.3) VM server. 
An IP-hashed load balancer divides traffic over eight separate instances of shiny-server, reducing the number of concurrent users per shiny-server.
We measure the performance of the model based on this infrastructure.

The most computationally demanding part of the care path model was written in the Julia programming language \cite{bezanson2017julia} to improve performance. 
Julia is rapidly gaining momentum as a tool for scientific computing due to higher performance compared to languages like R or Python without compromising ease of use.
The care path model written in Julia takes approximately 1 hundredth of the processing time of the R equivalent. 
We use JuliaConnectoR (v.1.0.0.9009) \cite{lenz_juliaconnector_2022} to allow communication between the Julia and R code bases.
One of the drawbacks of Julia is that the functions require just-in-time compilation the first time they are used;
the overhead related to compilation time exceeds the running time of the entire care path model.
To address this issue, we generated a fully compiled version of the model code \cite{PackageCompiler}, virtually eliminating loading time.
Once the main server is started, a concurrent Julia session is created and is shared between all shiny-server sessions.

\subsection{User interface}
The user interface of the dashboard is divided into sections (tabs) following the main modules of the model,
in combination with a side panel showing the basic controls: the catchment area selection, the forecast start date, the number of simulation runs, and the simulation length in days.
On each tab, parameter choices can be made that affect the current and further tabs, but not any previous tabs.
In this way, we prevent unnecessary re-running of the simulations.

Tabs are ordered as follows: 
1) Incidence, showing the daily reported number of cases with and without the nowcast. 
2) Vaccination, showing the cumulative number of vaccine doses administered, as well as the forecast of vaccinations.
It also includes the option to change the assumptions on the future vaccinations (number of first doses and booster doses per day, and the minimum delay between 2nd dose and booster). 
3) Effective R, showing the time-varying R estimation based on the EpiEstim package. 
This tab also includes the option to use the ETS model and the option to include a variant of concern (VoC), together with the needed VoC parameters.
4) Incidence forecast, showing the results of the incidence model, with the option to include the reporting pattern related to the weekdays. 
5) Bed forecast, reporting the results of the within-hospital care path model.  
The model is run using data on the current number of occupied beds in the hospital of interest, inputted either manually or by uploading a specifically structured file. 
6) Parameter choices. This tab includes all parameters included as default values in the other tabs, which should only be changed by users with more advanced knowledge of the underlying model.

\begin{figure}[H]
\begin{center}
\includegraphics[width=0.9\textwidth]{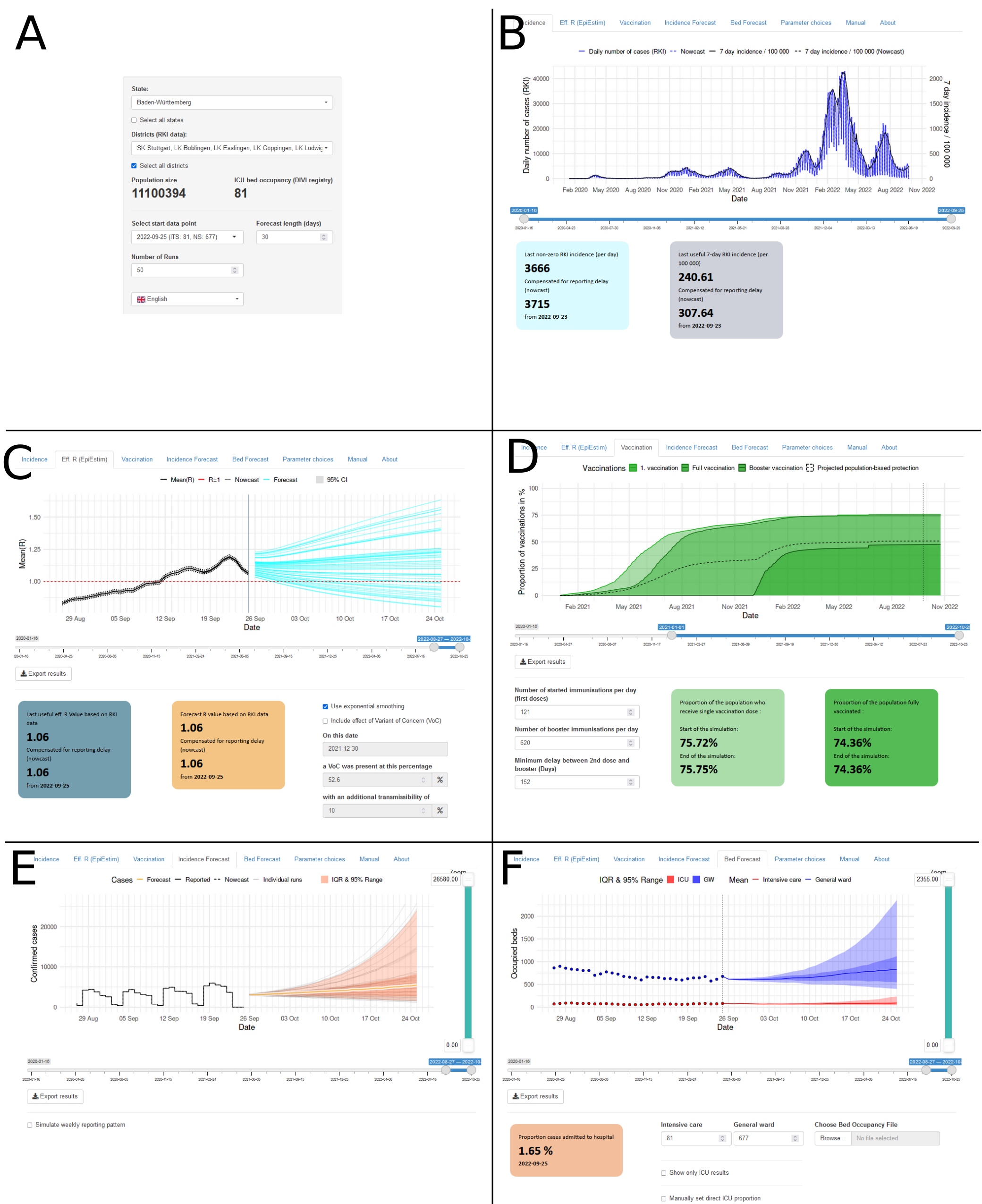}
\caption{The user interface of the on-request COVID-19 bed demand forecasting model. A) Side panel with basic controls, B) Reported incidence, C) Effective R forecast, D) Vaccination forecast, E) Incidence forecast, and F) Bed occupancy forecast.}
\label{fig:tabs}
\end{center}
\end{figure}

\section{Results}
\subsection{Performance}
Between 23-12-2021 and 07-04-2022, the online dashboard was visited 2352 times (counted as the number of unique sessions) by 575 users (counted as the number of unique machines by cookie UUID).
On average, sessions lasted 22 minutes, but the distribution of session durations is heavily right-skewed, ranging between 0 seconds (i.e., immediately closed by the user) and 12 hours, and a median of 1.6 minutes (IQR 0.3 - 5.9 minutes).
There were active sessions during 471 hours of the 2523 hours.

For 219 hours, the dashboard served multiple concurrent users, specifically more than three users for 44.5 hours, to a maximum of six users on 23-12-2021 and 19-01-2022.

\begin{figure}[htb]
\begin{center}
\includegraphics[width=\textwidth]{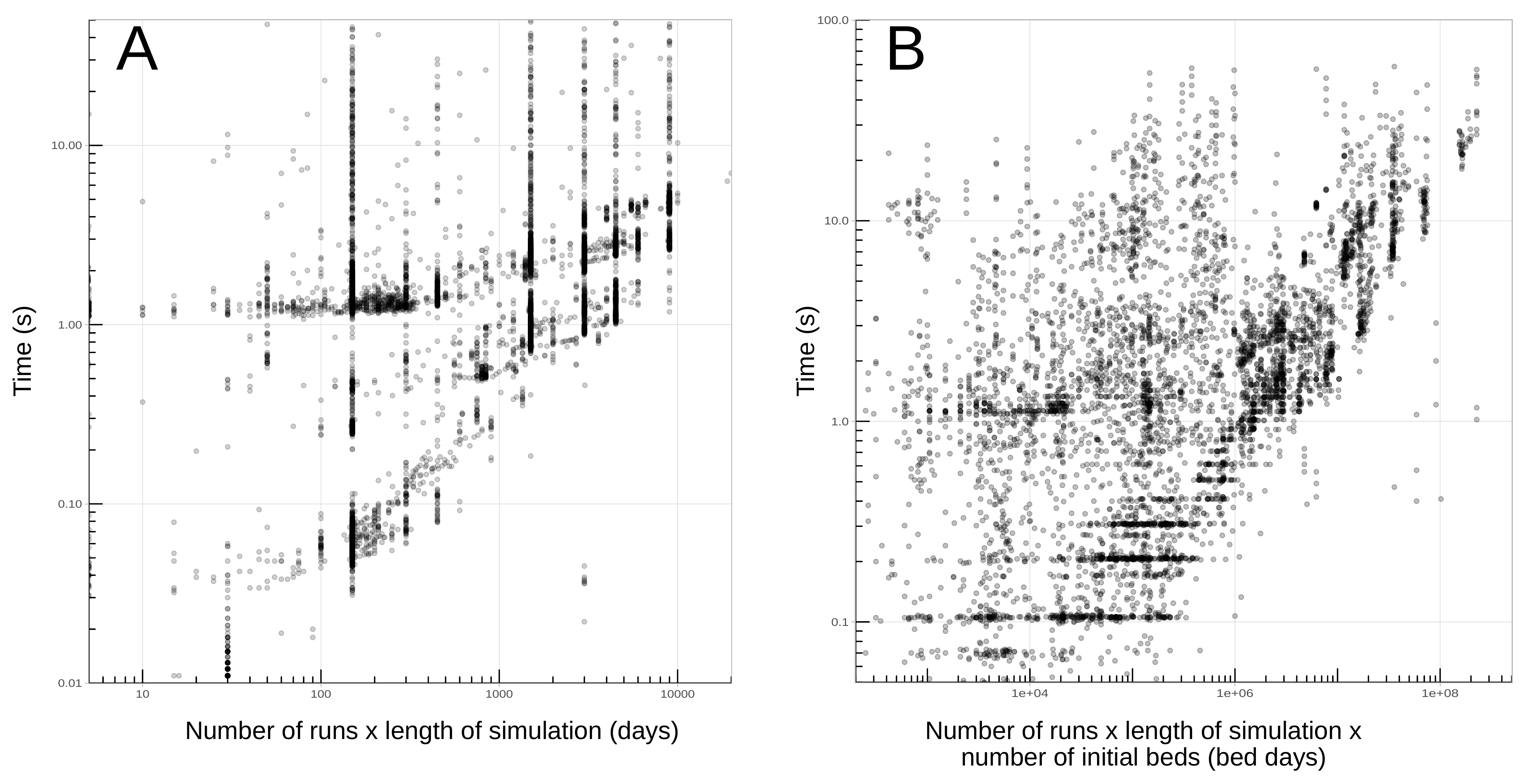}
\caption{The run-time of B) the incidence model and B) the care path model, as a function of the total number of run-days and bed-run-days, respectively.}
\label{fig:runtime}
\end{center}
\end{figure}

Users primarily selected hospital catchments consisting of a single district (n=2507), 
followed by 44 districts (N=1426), coinciding with the number of districts in the state of Baden-Württemberg, 
and 412 districts (N=807), indicating all districts in Germany.  
The incidence model was called 6875 times and took 2.7s on average (median 1.4s IQR 0.8-2.5s),
with a strong dependency on the total number of run-days (number of runs $\times$ length of the simulation, see figure \ref{fig:runtime}).
The care path model was called 5735 times and took 4.2s on average (median 1.9s IQR 0.9-4.1s).

As with the incidence model, the run time of the care path model depended strongly on the number of run-days.
Additionally, it depends on the total number of beds occupied at the start of the simulation.
On 157 occasions, users uploaded a file specifying the number of beds occupied in a specific hospital.
This uploading was done by 24 unique users, of which two were responsible for the vast majority of uploading instances (74 and 33 times).
The other forecasts were thus based on the number of beds occupied in each of the districts as automatically loaded from the central database.

\section{Conclusion}
This work shows that hospital-specific on-request production of bed demand forecasts during pandemics is possible.
The developed on-request forecasting tool offers users maximum flexibility by allowing the definition of hospital-specific assumptions and/or starting conditions, while at the same time basing all forecasts on the same model.
The combination of user-defined inputs within an existing model framework allows public health experts to adapt advanced forecasting methods to their local setting without having to build such models from scratch. 
Such an approach has the potential to make mathematical modelling of infectious diseases available to a much larger group of stakeholders, each with their specific set of questions to be answered.

Although our dashboard had a relatively modest number of users during the study duration, we are confident that it can cope with many more users, given the low proportion of time concurrent users were present. 
Nonetheless, there is certainly a limit to the traffic capacity of the dashboard, and particularly peak demand may be challenging.
For this reason, we avoided publicly promoting our dashboard, relying instead on our network of hospitals and other healthcare organisations to circulate it.
However, the dashboard was openly and freely available to any user over the entire period.
Furthermore. the code is freely available and can be installed and run locally if hospitals need their own instance of the dashboard.

In conclusion, on-request forecasts are a fundamentally different method of providing infectious disease forecasts compared to the more common static reports.
We showed how our dashboard implementation solves the specific set of challenges related to open-ended dynamic forecasting platforms, in particular computational requirements. 
The dashboard was successfully used by local healthcare providers, hospitals, and healthcare policymakers to evaluate incidence and hospital bed occupancy in Germany during the 2020-2022 COVID-19 pandemic.
We argue that these on-request forecasts are much more helpful in informing stakeholders at a local level where health management decisions, such as cancelling elective surgeries, directly affect the bed capacity. 
This way, the pandemic or epidemic response can be driven in near real-time on the level where it matters most.

\section*{Acknowledgements}
We thank Prof Dr Martin Wolkewitz, Dr Paul Biever, Dr Johannes Kalbhenn, Prof Dr Hartmut Bürkle, Prof Dr Winfried Kern, and Prof Dr Frederik Wenz for input about the parameters of the care paths model.

\section*{Conflicts of Interest}
Angelo D'Ambrosio is employed at the European Centre for Disease Prevention and Control at the time of pubblication but not at the time of drafting this manuscript. The views and opinions expressed herein are the authors’ own and do not necessarily state or reflect those of ECDC. ECDC is not responsible for the data and information collation and analysis and cannot be held liable for conclusions or opinions drawn.
The other authors have no conflicts of interest to declare.

\bibliography{DashboardTech}

\newpage
\section*{Supplementary material}
\subsection{Variables and parameters}
\begin{tabularx}{\textwidth}{lllX}
&Variable &Source&Description\\
\hline
\multirow{4}{*}{Catchment}&$K_{all}$ &Database&Set of all possible districts\\
&$K_{u}$ &User&Set of selected districts\\
&$N_g$ &Database&Population size of district $g$\\
&$N$ &Database&Total population size of all selected districts\\

\hline
\multirow{8}{*}{Incidence}&$I_g(t)$&Database&Number of infected individuals reported on day $t$ in district $g$\\
&$I(t)$ &Derived&Number of infected individuals reported on day $t$ in all selected districts\\
&$I_B(t)$&Derived&Number of individuals infected with the background variant on day $t$ in all selected districts\\
&$I_V(t)$&Derived&Number of individuals infected with the variant of concern on day $t$ in all selected districts\\

\hline
\multirow{7}{*}{Susceptibles}&$S(t)$&Derived&Total number of susceptible individuals at day $t$\\
&$S_B(t)$&Derived&Number of individuals susceptible to infection with the background variant on day $t$ in all selected districts\\
&$S_V(t)$&Derived&Number of individuals susceptible to infection with the variant of concern on day $t$ in all selected districts\\

\hline
\multirow{13}{*}{Vaccination}&$V_{d,g}(t)$ &Database&Number of $d^{th}$ doses administered on day $t$ in district $g$\\
&$V_d(t)$ &Derived&Number of $d^{th}$ doses administered on day $t$ in all selected districts\\
&$E_d$&User&Vaccine efficacy of the $d^{th}$ dose (Default: Dose 1: 25\% dose 2: 50\%, dose 3: 78\%)\\
&$G_d(t)$&User&Delay function for vaccine effect. Proportion of cases on day $t$ after administration of dose $d$ protected to the maximum $E_d$. (Default: see SI)\\
&$G_P(t)$&Derived&Population vaccine-based protection against infection, proportion of the population not susceptible to infection\\
&$\epsilon$&User&Immune evasion\\

\hline
\multirow{21}{*}{Transmission}&$R_e(t)$&Derived&Time-varying effective reproduction number at day $t$\\
&$R_0(t)$&Derived&Basic reproduction number at day $t$\\
&$R_{0,B}(t)$&Derived&Basic reproduction number of the background viral population at day $t$\\
&$R_{0,V}(t)$&Derived&Basic reproduction number of the variant of concern at day $t$\\
&$\sigma(t)$&User&Serial interval distribution, giving the proportion of new cases arising $t$ days after their infector (Default: 5 days)\\
&$\beta(t)$&Derived&Transmission rate\\
&$P(t)$&Derived&Infection pressure; Total number of past cases weighted by the serial interval\\
&$P_I(t)$&Derived&The infection pressure exerted on each individual within the population\\

&$A$&User&Added fitness advantage of the variant of concern\\
&$A_\epsilon$&Derived&Added fitness advantage of the variant of concern, through immune evasion \\
&$\rho_B(t)$&Derived&Backround viral population proportion among all cases\\
&$\rho_V(t)$&Derived&Variant of concern proportion among all cases\\
&$r$&Derived&Growth rate\\

\hline
\multirow{8}{*}{Bed occupancy}&$B_{GW,g}(t)$ &Database&Number general ward beds occupied on day $t$ in district $g$ \\
&$B_{ICU,g}(t)$ &Database&Number intensive care beds occupied on day $t$ in district $g$\\
&$B_{GW}(t)$ &Derived /&Number general ward beds occupied on day $t$\\
& &User&\\
&$B_{ICU}(t)$ &Derived /&Number intensive care beds occupied on day $t$\\
& &User&\\

\hline
\multirow{30}{*}{Care path model}&$H_I(t)$&Derived&Proportion of all patients admitted to hospital directly admitted to the ICU\\
&$L(t)$&Derived&Proportion of cases still in hospital $t$ days after admission\\
&$L_{GW}(t)$&Derived&Proportion of cases present on the General ward $t$ days after admission to the hospital\\
&$L_{ICU}(t)$&Derived&Proportion of cases present on the Intensive Care Unit $t$ days after admission to the hospital\\

&$\alpha(t)$&Derived&Proportion of all cases admitted to hospital (Admission rate)\\
&$\alpha_{GW}(t)$&Derived&Proportion of all cases admitted to the General Ward (Admission rate)\\
&$\alpha_{ICU}(t)$&Derived&Proportion of all cases admitted to the ICU (Admission rate)\\
&$X_B(t)$&Derived&Number of infected cases with the background variant with pre-existing immunity\\
&$X_V(t)$&Derived&Number of infected cases with the Variant of Concern with pre-existing immunity\\
&$X^*(t)$&Derived&Proportion of all infected cases with pre-existing immunity \\
&$U(t)$&Derived&Proportion of all individuals that have immunity through previous infection and/or vaccination\\
&$U_B(t)$&Derived&Proportion of all individuals that have immunity against the background viral population through previous infection and/or vaccination\\
&$U_V(t)$&Derived&Proportion of all individuals that have immunity against the VoC through previous infection\\
&$\pi$&Derived&Primary infection hospitalisation risk\\
&$\gamma$&User&Hospitalisation risk relative to primary infection (Default 0.1)\\

\hline
\multirow{5}{*}{Time}&$T_S$&User&Start date of the simulation\\
&$\Delta T_V$&Derived&Time between first and second dose\\
&$\Delta T_B$&User&Time between second and third dose (Default: 152 days)\\
&$t_\rho$&User&Reference date for the Variant of Concern\\
\hline
\end{tabularx}

\end{document}